\documentclass[conference]{IEEEtran}
\IEEEoverridecommandlockouts
\usepackage{caption}
\usepackage{multirow}
\usepackage{booktabs} 
\usepackage{float}
\usepackage{array}
\usepackage{cite}
\usepackage{amsmath,amssymb,amsfonts}
\usepackage{algorithmic}
\usepackage{graphicx}
\usepackage{textcomp}
\usepackage{xcolor}
\usepackage{listings}
\usepackage{tcolorbox}
\usepackage{url}  % this before hyperref
\usepackage{hyperref}  % should come last
\hypersetup{
    colorlinks=true,
    allcolors=black
}

\definecolor{codegreen}{rgb}{0,0.6,0}
\definecolor{codegray}{rgb}{0.5,0.5,0.5}
\definecolor{codepurple}{rgb}{0.58,0,0.82}
\definecolor{backcolour}{rgb}{0.95,0.95,0.92}

\definecolor{zwcolor}{rgb}{0.5,0.7,0.9}

\lstdefinestyle{mystyle}{
    backgroundcolor=\color{backcolour},   
    commentstyle=\color{codegreen},
    keywordstyle=\color{magenta},
    numberstyle=\tiny\color{codegray},
    stringstyle=\color{codepurple},
    basicstyle=\ttfamily\footnotesize,
    breakatwhitespace=false,         
    breaklines=true,                 
    captionpos=b,                    
    keepspaces=true,                 
    numbers=left,                    
    numbersep=5pt,                  
    showspaces=false,                
    showstringspaces=false,
    showtabs=false,                  
    tabsize=2
}

\def\BibTeX{{\rm B\kern-.05em{\sc i\kern-.025em b}\kern-.08em
    T\kern-.1667em\lower.7ex\hbox{E}\kern-.125emX}}
\begin{document}

\title{Secret Breach Detection in Source Code with Large Language Models\\}

% \author{\IEEEauthorblockN{Anonymous Authors}
% \IEEEauthorblockA{\textit{}}} 

\author{\IEEEauthorblockN{Md Nafiu Rahman\IEEEauthorrefmark{1}}
\IEEEauthorblockA{\textit{Department of CSE} \\
\textit{Bangladesh University of Engineering and Technology}\\
Dhaka, Bangladesh \\
nafiu.rahman@gmail.com}
\and
\IEEEauthorblockN{Sadif Ahmed\IEEEauthorrefmark{1}}
\IEEEauthorblockA{\textit{Department of CSE} \\
\textit{Bangladesh University of Engineering and Technology}\\
Dhaka, Bangladesh \\
ahmedsadif67@gmail.com}
\and
\IEEEauthorblockN{Zahin Wahab\IEEEauthorrefmark{1}}
\IEEEauthorblockA{\textit{Department of CS} \\
\textit{The University of British Columbia}\\
Vancouver, BC, Canada \\
zahinwahab@gmail.com}
\and
\IEEEauthorblockN{S M Sohan}
\IEEEauthorblockA{\textit{Sr. Engineering Manager} \\
\textit{Google}\\
sohan39@gmail.com}
\and 
\IEEEauthorblockN{Rifat Shahriyar}
\IEEEauthorblockA{\textit{Department of CSE} \\
\textit{Bangladesh University of Engineering and Technology}\\
Dhaka, Bangladesh \\
rifat@cse.buet.ac.bd}

\thanks{\IEEEauthorrefmark{1}The first three authors contributed equally to this work.} \\

}

\IEEEpubid{\makebox[\columnwidth]{979-8-3315-9147-2/25/\$31.00~\copyright2025 IEEE\hfill} \hspace{\columnsep}\makebox[\columnwidth]{ }}

\maketitle

\IEEEpubidadjcol

\begin{abstract}
Background: Leaking sensitive information—such as API keys, tokens, and credentials—in source code remains a persistent security threat. Traditional regex and entropy-based tools often generate high false positives due to limited contextual understanding. Aims: This work aims to enhance secret detection in source code using large language models (LLMs), reducing false positives while maintaining high recall. We also evaluate the feasibility of using fine-tuned, smaller models for local deployment.
Method: We propose a hybrid approach combining regex-based candidate extraction with LLM-based classification. We evaluate pre-trained and fine-tuned variants of various Large Language Models on a benchmark dataset from 818 GitHub repositories. Various prompting strategies and efficient fine-tuning methods are employed for both binary and multiclass classification. Results: The fine-tuned LLaMA-3.1 8B model achieved an F1-score of 0.9852 in binary classification, outperforming regex-only baselines. For multiclass classification, Mistral-7B reached 0.982 accuracy. Fine-tuning significantly improved performance across all models. Conclusions: Fine-tuned LLMs offer an effective and scalable solution for secret detection, greatly reducing false positives. Open-source models provide a practical alternative to commercial APIs, enabling secure and cost-efficient deployment in development workflows.

\end{abstract}


\section{Introduction}
\begin{figure}[h]
    \centering
    \includegraphics[width=0.8\linewidth]{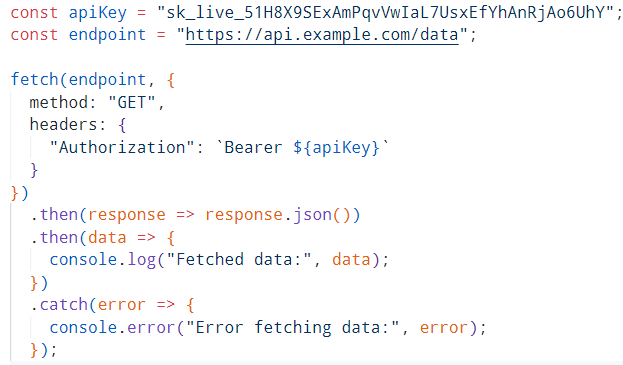}
    \caption{Example of a secret breach in source code: An API key is hardcoded and exposed, posing a security risk if the code is shared or pushed to a public repository.}
    \label{secretbreach-example}
\end{figure}

Secrets play a crucial role in software development by serving as a digital authentication mechanism that protects access to sensitive systems and services. These secrets take many forms, such as API tokens, OAuth credentials, encryption keys (notably RSA), TLS/SSL certificates, and authentication details like usernames and passwords \cite{gitguardian2025secrets}. However, as modern applications rely on interconnected services, secrets can unintentionally spread across different components, amplifying the risk of exposure \cite{secretsaboutsecretsincode}. Figure \ref{secretbreach-example} shows an example of secret breach in source code. The JavaScript code above illustrates the use of the \texttt{fetch} API to perform an authenticated HTTP \texttt{GET} request. The authorization is handled via a Bearer token included in the request header. The response is parsed as JSON and logged to the console. Any errors encountered during the request are captured and reported.

Platforms like GitHub and Bitbucket are vital for developer collaboration but often lead to inadvertent secret leaks. GitGuardian's reports show a sharp rise in such incidents: in 2022 alone, 10 million hardcoded secrets were detected—a 67\% increase from 2021—as scanned commits rose from 860 million to over 1 billion \cite{secretsprawlreport}. Exposed API keys have enabled attackers to exploit free services \cite{stealingAWScredentials}, and Sophos found that compromised credentials caused half of all cyberattacks in early 2023 \cite{sophosreport}. Despite detection efforts by cloud providers, version control systems provide limited protection. Leaks occur not only in source code but also in config files, test scripts, docs, dependency files, and logs \cite{secretsaboutsecretsincode}.
The 2024 report \cite{secretsprawlreport24} highlighted a 1212× surge in leaked OpenAI API keys—the fastest-growing secret type—at a rate of 46,000 per month. Particularly alarming are “zombie leaks,” which remain valid even after repos are deleted or privatized, posing long-term risks. \newline
A variety of tools \cite{secretdetectiontools} exist to identify vulnerabilities during code commits, including TruffleHog \cite{trufflehog2}, Gitrob \cite{gitrob}, ggshield \cite{gitguardian}, and Gitleaks \cite{gitleaks}. These tools typically rely on regex-based \cite{trufflehogregexes} pattern matching and entropy analysis. For instance, Gitleaks and ggshield use predefined regex patterns to detect hardcoded secrets, whereas TruffleHog identifies high-entropy strings \cite{shannon1948mathematical}. However, such techniques often yield high false positive rates \cite{basak2023comparative}, making them less effective in noisy environments.\newline
To address these limitations, Meli et al. \cite{meli2019bad} conducted a comprehensive study on GitHub repositories, combining pattern matching with entropy-based filtering via regex. Their approach successfully detected secrets in over 100,000 repositories with an accuracy of 99.29\%. Further advancements were made by Sinha et al. \cite{sinha2015detecting}, who incorporated keyword searches and program slicing techniques to improve both recall and precision in identifying API keys and similar credentials. Meanwhile, Saha et al. \cite{saha2020secrets} tackled false positives by training machine learning models on datasets retrieved through the Git REST API \cite{restAPI}, achieving an F1-score of 86.7\% using a Soft-Voting Classifier. Feng et al. \cite{feng2022automated} further refined secret detection by applying deep neural networks to reduce false positives in textual password detection. To mitigate the risks associated with secret leaks, Sinha et al. \cite{sinha2015detecting} suggested various preventive measures, including the use of tools like \texttt{git-crypt} \cite{gitcrypt}, which encrypts sensitive data in Git repositories.
Wahab et al. \cite{wahab2024secretbreachpreventionsoftware} used transformer models like BERT \cite{devlin2019bert} to detect secrets in Github issue reports. 
% Additionally, Basak et al. \cite{basak2022practices} compiled 24 best practices for effective secret management, offering structured guidelines to minimize exposure risks

In this work, we propose a novel approach to automate secret detection in source code by leveraging large language models (LLMs). Our methodology begins with regex-based candidate string extraction, after which an LLM classifies true secrets, significantly reducing false positives. We explore several strategies, including zero-shot inference both before and after fine-tuning, and one-shot and few-shot prompting on pre-trained models.. Our fine-tuned model achieves an F1-score of 0.9852 on the test dataset, outperforming regex-based and conventional machine learning approaches.

Our key contributions are as follows:

\begin{itemize}
    \item We develop a hybrid methodology that first uses regex-based candidate extraction and then refines classification using LLMs, reducing false positives while maintaining high recall.
    \item We test multiple LLM-based detection strategies, including zero-shot, one-shot, few-shot learning, and fine-tuned inference, to assess their impact on precision and recall.
    \item We extend the detection approach to multiclass classification, categorizing secrets into types such as Private Keys, API Keys, Authentication Tokens, and more, significantly enhancing the granularity of detection.
\end{itemize}

To this end, we aim to address the following research questions:

\begin{itemize}
    \item \textbf{RQ1:} How effective are large language models in detecting secrets?
    \item \textbf{RQ2:} How effective are smaller models for local server deployment?
\end{itemize}

The rest of this paper is organized as follows. Section~\ref{sec:related} reviews related work. Section~\ref{sec:methodology} describes the methodology, including data collection, preprocessing and data distribution. Section~\ref{sec:llm_detection} presents our LLM-based detection framework, covering prompting strategies and performance across different models. Section~\ref{sec:small_llm_detection} discusses the use of fine-tuned smaller LLMs for local deployment. Section~\ref{sec:discussion} provides a detailed discussion on evaluation insights, limitations, and impact factors. Section~\ref{sec:threats} outlines potential threats to validity, and finally, Section~\ref{sec:conclusion} concludes the paper and discusses future directions.

\textbf{Replication Package.} Our code and data are shared at 
\url{https://doi.org/10.5281/zenodo.15273217}

\section{Related Work}
\label{sec:related}

\paragraph{Large-Scale Secret Leak Analysis}
Meli et al.~\cite{meli2019bad} conducted a large-scale study on secret leaks in GitHub by scanning billions of files through real-time and snapshot-based analysis. Their regex and entropy-based filtering approach achieved high precision (99.29\%) but relied on a relatively small set of patterns and minimal manual labeling, potentially limiting coverage and generalizability. In contrast, our work uses a curated benchmark dataset with 768 regex patterns and 15,000 manually labeled secrets, leveraging large language models (LLMs) to incorporate contextual understanding.

\paragraph{Program Analysis for Secret Detection}
Building upon earlier approaches, Sinha et al.~\cite{sinha2015detecting} enhanced secret detection by incorporating program slicing, keyword-based, and pattern-based searches. The approach achieved up to 100\% precision and 84\% recall for Amazon AWS keys by leveraging static analysis techniques. While impactful, the method is useful for specific patterns and language constructs, limiting its adaptability across diverse codebases.

\paragraph{Secret Detection Tools}
Tools like TruffleHog and Gitleaks scan repositories for hardcoded secrets using regexes, entropy checks, and known patterns. While effective for common cases, they often struggle with false positives and lack contextual understanding, especially for obfuscated or non-standard secrets.

\paragraph{Machine Learning-Based Filtering}
Saha et al.~\cite{saha2020secrets} used regex-extracted data and trained a Soft Voting Classifier on binary features like entropy and character patterns, achieving an F1-score of 86.7\%. However, their reliance on handcrafted features and lack of contextual understanding limited their performance. We address this by leveraging LLMs that capture semantic and structural cues from code, yielding significantly higher accuracy.

\paragraph{Deep Learning Approaches}
Feng et al.~\cite{feng2022automated} applied deep neural networks to detect textual passwords, reducing false positives through semantic learning. Yet, their focus was narrow and lacked generalization across diverse secret types. Our LLM-based method extends deep learning to diverse secrets in code, supporting both binary and multiclass classification.

\paragraph{Secret Detection in Issue Reports}
Wahab et al.~\cite{wahab2024secretbreachpreventionsoftware} focused on secret detection within issue reports and documentation using transformer-based language models such as BERT and RoBERTa. Their results highlighted the benefits of contextual embeddings in improving detection accuracy for non-code textual software artifacts.

% \paragraph{Secret Mitigation Techniques}
% In terms of mitigation, Sinha et al.~\cite{sinha2015detecting2} suggested strategies like excluding secrets from version control, encrypting them using tools such as \textit{git-crypt}, and integrating automated scanning mechanisms. They also emphasized practices like commit signing and two-factor authentication (2FA) to strengthen security.

% \paragraph{Best Practices for Secret Management}
% Best practices for secret management were compiled by Basak et al.~\cite{basak2022practices}, who proposed 24 actionable guidelines. These included avoiding hard-coded secrets, using environment variables, employing short-lived credentials, utilizing secret managers like Vault, and sanitizing version control system (VCS) history.

\paragraph{Comparative Analysis of Detection Tools}
Finally, Basak et al.~\cite{basak2023comparative} conducted a comparative analysis of both open-source and proprietary secret detection tools. They evaluated tools based on categorization, scoring methods, and detection capabilities, providing insights to help practitioners choose solutions based on precision, recall, and usability.

While prior work has predominantly focused on regex-based or machine learning-based techniques for detecting secrets in textual artifacts, our research shifts the focus to using large language models (LLMs) for identifying secrets directly within source code. By fine-tuning these models, we significantly reduce false positives and enhance detection accuracy, effectively bridging the gap between traditional approaches and modern deep learning within real-world development workflows.

\section{Methodology}
\label{sec:methodology}
\subsection{Data Collection and Preprocessing}
\label{sec:methodology}
\subsubsection{Data Collection}
Our dataset is derived from \textbf{SecretBench} \cite{basak2023secretbench2}, a publicly available benchmark dataset containing \textbf{97,479 candidate secrets} extracted from \textbf{818 public GitHub repositories}. Among these, \textbf{15,084 are labeled as true secrets}, manually verified by analyzing both the secret strings and their surrounding code context. The dataset spans \textbf{49 programming languages} and \textbf{311 different file types}, covering various forms of secrets such as API keys, OAuth tokens, cryptographic private keys, and database credentials.

The original SecretBench dataset was constructed using \textbf{761 predefined regular expressions (regex)}\cite{secretbenchregexes} to identify potential secrets from GitHub repositories. These regex patterns targeted a diverse range of secret types, including high-entropy strings, access tokens, and password patterns. Each detected candidate was manually labeled to distinguish between actual secrets and false positives, such as dummy credentials or placeholders. The dataset also includes metadata such as file paths, commit IDs, entropy scores, and programming language information.

\subsubsection{Preprocessing}
To prepare the dataset for training and evaluation, we performed multiple preprocessing steps:

\begin{enumerate}
    \item \textbf{File Content Extraction} – Each candidate secret's corresponding source file was retrieved from the \texttt{Files.zip} dataset. File contents were read while handling encoding variations, with unreadable characters removed to ensure consistency.
    
    % \item \textbf{Noise Reduction} – Certain types of content, such as \textbf{log files, URLs, commit IDs, stack traces, shell commands, UUIDs, and dummy passwords}, were filtered out to minimize false positives.

    \item \textbf{Dataset Balancing} – Since the dataset is highly imbalanced, we curated a subset for training and evaluation:
    \begin{itemize}
        \item \textbf{30,000 non-secret samples} were randomly selected.
        \item \textbf{15,000 true secret samples} were retained.
    \end{itemize}
  
\end{enumerate}

For multiclass classification, we only took the positive samples from the dataset. The dataset distribution is mentioned in table \ref{multi-table dist}. The \textit{Other} category includes various forms of sensitive values that do not fall under standard API keys or credentials. These may include a Package Key ID (e.g., PKG-8736-XYZ), an Instrumentation Key (e.g., a1b2c3d4-e5f6-4711-9a8b-112233445566), or a random string used for internal tracking or authentication purposes.

\begin{table}[h]
\centering
\caption{Multiclass Secret Distribution (15,084 Samples)}
\label{multi-table dist}
\begin{tabular}{|l|c|}
\hline
\textbf{Secret Type} & \textbf{Count} \\
\hline
Private Key                     & 5,791 \\
API Key and Secret             & 4,521 \\
Authentication Key and Token  & 3,567 \\
Other                          & 524   \\
Generic Secret                 & 333   \\
Database and Server URL       & 162   \\
Password                       & 150   \\
Username                       & 27    \\
\hline
\end{tabular}
\vspace{-1em}
\end{table}

To train our LLMs on multiclass secret detection, we split the labeled dataset as follows:

\begin{itemize}
    \item Training Set: 9,000 samples
    \item Evaluation Set: 3,000 samples
    \item Test Set: 3,000 samples
\end{itemize}

We followed the same preprocessing steps as in binary classification.

\begin{figure}[h]
    \centering
    \includegraphics[width=1\linewidth]{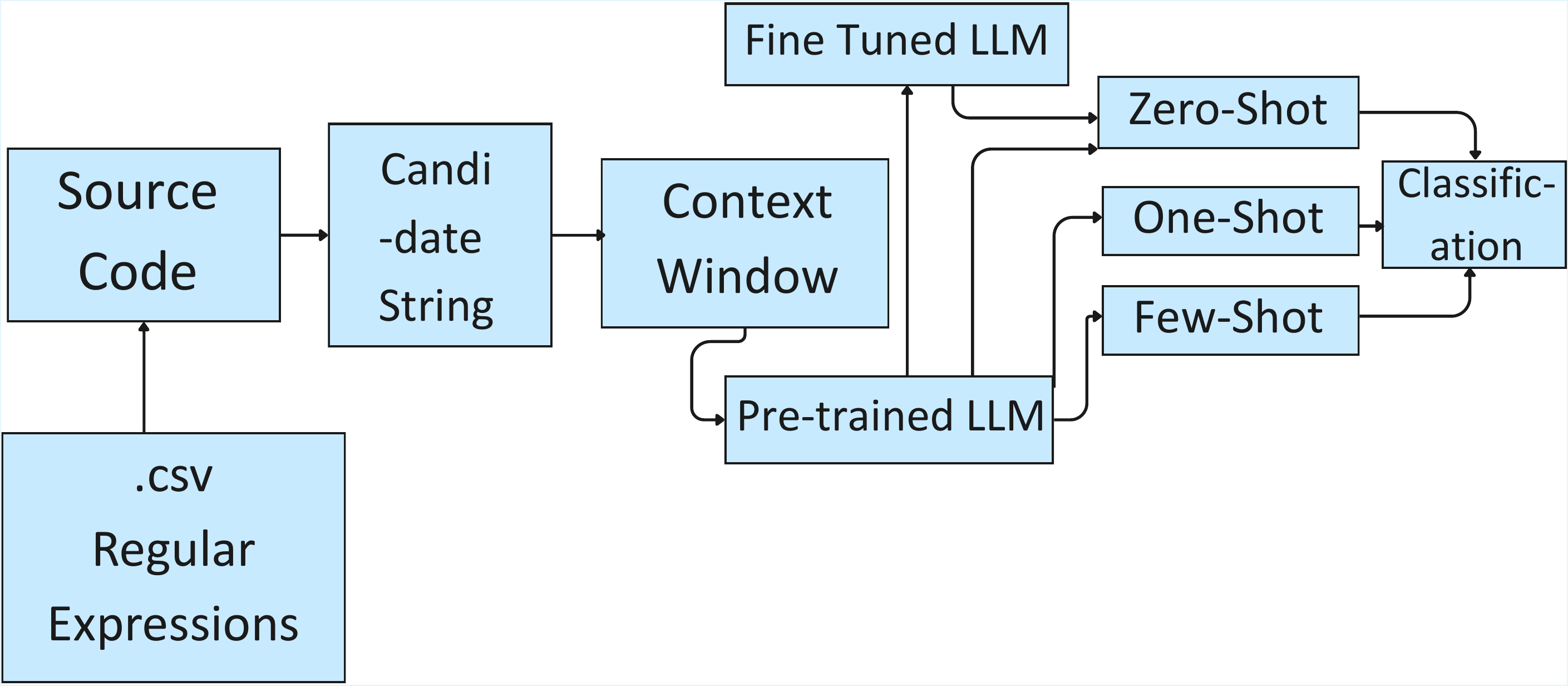} 
    \captionsetup{font=normalsize}
    \caption{Workflow for secret detection in source code. Potential candidate strings are extracted using 761 regular expressions from a .csv file and located within the source code. A 200-character context window around each candidate is then created. This context is either fed directly into a pretrained LLM for zero-shot, one-shot, or few-shot classification, or passed through a fine-tuned LLM (trained on the SecretBench dataset) for downstream zero-shot classification.}

    \label{pipeline}
\end{figure}

Figure~\ref{pipeline} illustrates the overall workflow of our proposed secret detection system. The process begins with the extraction of candidate strings using a set of predefined regular expressions. For each candidate, we construct a contextual window from the surrounding source code to provide the necessary semantic cues. These context-enriched inputs are then classified using either pre-trained or fine-tuned large language models (LLMs). The figure serves as a high-level guide to the methodology, with each component discussed in detail in the following sub-sections.

\subsection{Secret Detection Using Pre-trained Large Language Models (RQ1)}
\label{sec:llm_detection}
\subsubsection{Motivation}
Pre-trained large language models (LLMs) have shown remarkable performance have demonstrated exceptional capabilities in text classification \cite{kostina2025largelanguagemodelstext} and contextual understanding \cite{zhu2024largelanguagemodelsunderstand}, making them well-suited for secret detection. Their ability to capture contexts in code \cite{nguyen2024empiricalstudycapabilitylarge}, even without extensive task-specific fine-tuning, makes them promising for security-related applications. Motivated by this, we investigate the inherent capability of LLMs to identify secrets embedded in source code. Specifically, we explore how well these models can distinguish true secrets from non-sensitive strings using contextual understanding, even without extensive retraining.

\subsubsection{Approach}
In this study, we evaluated GPT-4o \cite{achiam2023gpt} and Deepseek-V3 \cite{liu2024deepseek} on detecting secrets as and experimented with different prompting strategies such as zero-shot, one-shot, and few-shot learning.

\textbf{Context Extraction and Data Preprocessing: }  
Secret detection often requires an understanding of the surrounding context \cite{contextwindow} \cite{cwd2001} in which a candidate string appears. To provide sufficient contextual information, we extract a \textbf{200-character window} around each candidate secret from its source code. This allows the model to better differentiate between sensitive and non-sensitive patterns. Moreover, since source code files can span thousands of lines, feeding the entire file into a model is computationally expensive and often impractical.

Each data point consists of:  
\begin{itemize}
    \item \textbf{Candidate String}: The potential secret identified from the dataset.
    \item \textbf{Code Snippet}: The extracted surrounding code providing contextual information.
    \item \textbf{Label}: A binary classification label indicating whether the candidate is a \textit{Secret} or \textit{Non-sensitive}.
\end{itemize}
The extracted data is then formatted into structured natural language prompts to facilitate LLM-based classification.

\textbf{Prompt Engineering:} The interaction with large language models (LLMs) is primarily conducted through prompt engineering, where a carefully crafted textual input is provided to the model to perform a specific task. In our case, we prompt the model to classify whether a given code snippet contains a secret or not. This is achieved by presenting the model with candidate strings and their surrounding context in source code. We explore several prompting strategies, including zero-shot, one-shot, and few-shot prompting, inspired by prior work on LLM-driven generation tasks such as test case creation and code understanding.

\begin{figure}[h]
    \centering
    \includegraphics[width=1\linewidth]{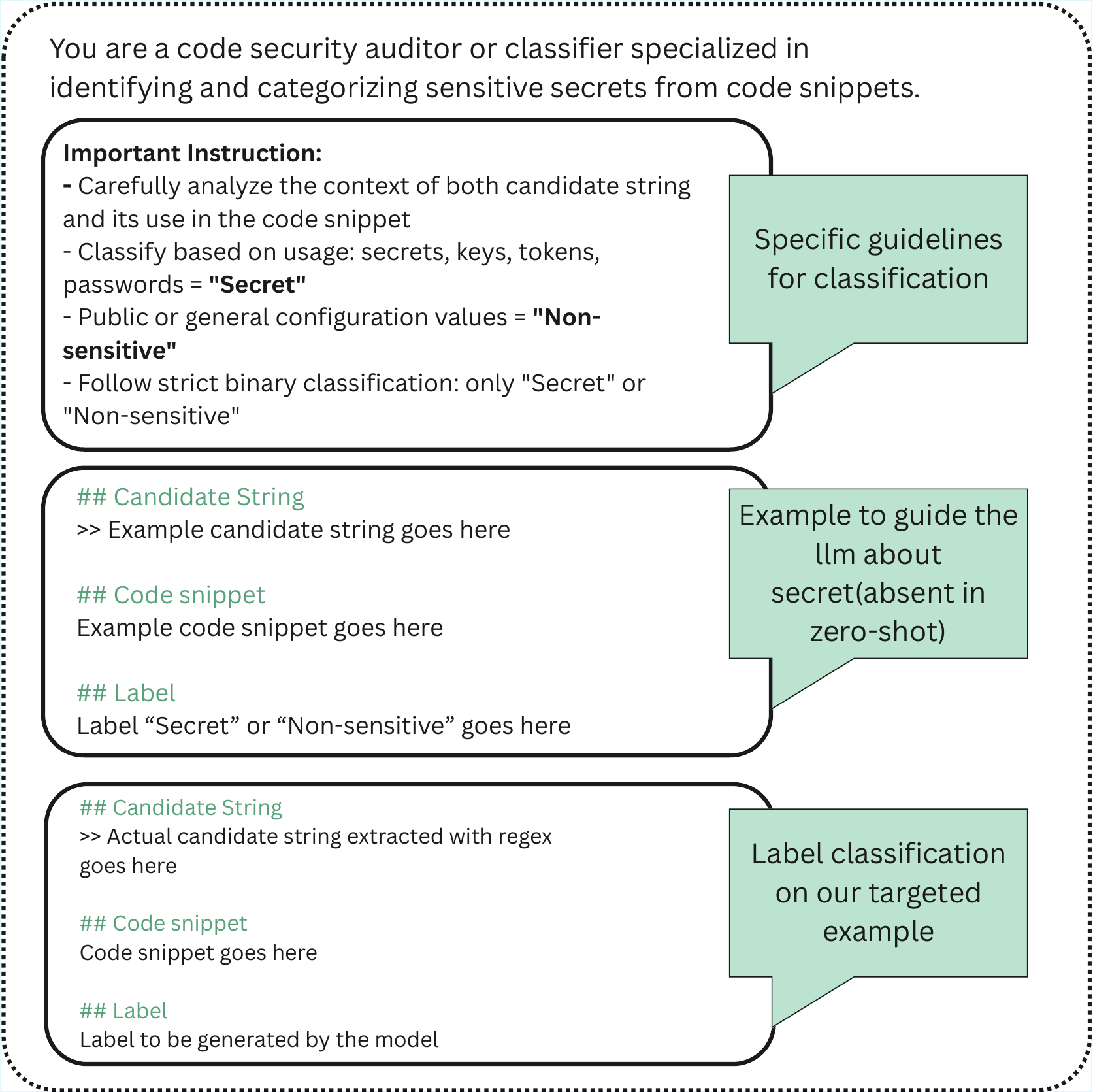}
    \caption{Prompt for classifying a candidate string as "Secret" or "Non-sensitive" from a code snippet using pre-trained LLMs.}
    \label{prompt-engineering}
\end{figure}

\paragraph{Zero-Shot Prompting}  
In the zero-shot setting, the model receives only the candidate string and its surrounding code snippet, without any labeled examples or instructions. It relies solely on its pre-trained knowledge to determine whether the string is a secret. This setting serves as a baseline to assess the model’s inherent reasoning ability. The prompt format used is shown in Figure~\ref{prompt-engineering}.

\paragraph{One-Shot Prompting}  
In one-shot prompting, a single labeled example with its code context is provided before the prediction task. This example demonstrates the classification process and guides the model’s decision-making. We select a general yet representative sample capturing common secret patterns.

\paragraph{Few-Shot Prompting}  
Few-shot prompting \cite{brown2020language} extends the one-shot setting by including multiple labeled examples. These cover varied secret and non-secret cases, offering richer context. This setup improves model performance, especially on ambiguous inputs, by allowing it to learn from analogical reasoning.

\textbf{Parameter Settings: } To control the behavior of LLM outputs during inference, we adjust key generation parameters. The \texttt{temperature} parameter controls the randomness of the output distribution. Lower values make the model more deterministic, while higher values introduce more variability. We use a low \texttt{temperature} value (set to 0.0) to reduce randomness and ensure deterministic outputs.

\textbf{Evaluation Data: }
For evaluation, we randomly selected a subset of 3,000 (1,500 secret + 1,500 non-secret) data points from our dataset.

\subsubsection{Multiclass Classification}
As mentioned in Section \ref{sec:methodology} we  extended our approach to multiclass classification. We evaluated using 3,000 test samples from the partition mentioned in the dataset.

\textbf{Evaluation Metrics:}  
To assess the performance of our LLM-based secret detection models, we employ standard classification evaluation metrics: precision, recall, F1-score, and F2-score. These metrics provide insights into how well the model distinguishes between secrets and non-sensitive strings.

\begin{itemize}
    \item \textbf{Precision:}  
    Precision measures how many of the instances predicted as secrets are actually true secrets. It reflects the proportion of true positives among all predicted positives. A high precision score indicates fewer false alarms, reducing unnecessary security alerts.

    \item \textbf{Recall:}  
    Recall measures how many actual secrets were correctly identified by the model. It captures the proportion of true positives among all actual positives. A high recall score ensures that most secrets are identified, reducing the risk of sensitive data exposure.

    \item \textbf{F1-Score:}  
    The F1-score is the harmonic mean of precision and recall. It provides a balanced performance measure, especially useful when both false positives and false negatives need to be minimized.

    \item \textbf{F2-Score:}  
    The F2-score is a weighted harmonic mean of precision and recall, placing more emphasis on recall. This is particularly relevant in security-sensitive applications, where failing to detect a secret (false negative) can have more serious consequences than a false positive.
    \begin{equation}    
    F_2 = \frac{(1 + 2^2) \cdot \text{Precision} \cdot \text{Recall}}{2^2 \cdot \text{Precision} + \text{Recall}}
    \end{equation}

\end{itemize}

\textbf{F1 and F2 Scores for Balancing Precision and Recall:}  While our dataset is balanced (12,000 secrets and 12,000 non-secrets), real-world secret detection tasks often prioritize capturing as many true secrets as possible. Accuracy is a reasonable metric during training, but it fails to reflect the trade-offs between false positives and false negatives. To evaluate models more effectively, we adopt both the F1-score and F2-score. The F1-score provides a balanced view, minimizing both types of error, while the F2-score emphasizes recall to reduce the likelihood of missed secrets. This dual-metric approach ensures that our evaluation aligns with the practical needs of security workflows—where over-detection can cause alert fatigue, and under-detection can lead to significant data breaches.

\subsection{Secret Detection with Smaller Fine-tuned Models (RQ2)}
\label{sec:small_llm_detection}
\subsubsection{Motivation} 
Many organizations are reluctant to adopt third-party large language models (LLMs) for software development tasks due to internal compliance policies, data privacy concerns, and the high costs associated with commercial APIs. To address these issues, we investigate the feasibility of using smaller, open-source models that can be fine-tuned and deployed on local servers. Models such as LLaMA \cite{touvron2023llamaopenefficientfoundation}, Gemma \cite{team2024gemma}, CodeLLaMA \cite{rozière2024codellamaopenfoundation}, and DeepSeek offer cost-effective alternatives with promising capabilities. When fine-tuned with high-quality data, these models often achieve performance levels comparable to larger general-purpose LLMs, while consuming significantly less memory and compute resources. This makes them ideal for deployment in constrained environments. Moreover, locally hosted models eliminate usage-based costs and ensure complete control over sensitive and proprietary data, enhancing both security and compliance.

\subsubsection{Approach} 
We fine-tuned several compact models— LLaMA-3.1-8B, Mistral-7B, CodeLLaMA-7B, DeepSeek-7B, and Gemma-7B to detect secrets. These models were selected for their strong performance on complex language tasks, efficient inference capabilities, and suitability for long-sequence processing, all while maintaining a low computational footprint. In addition to evaluating the fine-tuned models by zero-shot prompting on them \cite{wei2022finetunedlanguagemodelszeroshot}, we also assessed the raw, pre-trained versions to understand the benefits of fine-tuning. For the baseline comparison, we conducted zero-shot, one-shot, and few-shot prompting experiments on the pre-trained models without any task-specific training.

\textbf{Context Extraction and Data Preprocessing: }
We used the same context window approach like section III-B.

\textbf{Prompt Engineering:}
We used the same prompt format outlined in Section III-B for the fine-tuned models. We applied zero shot, one-shot and few-shot learning in raw pre-trained models without any task-specific training and applied zero-shot learning in fine-tuned models. 

\textbf{Parameter Settings: } 
For fine-tuning, we used \textbf{QLoRA} \cite{dettmers2023qloraefficientfinetuningquantized} with 4-bit precision (NF4 quantization and FP16 computation), disabling double quantization to reduce memory usage. Fine-tuning focused on transformer attention layers for domain-specific pattern learning. We employed parameter-efficient fine-tuning (PEFT) \cite{xu2023parameterefficientfinetuningmethodspretrained} using LoRA with rank 64 and $\alpha=16$, without dropout or bias adjustments. The \texttt{Paged AdamW} optimizer (learning rate $2\times10^{-4}$) was used with memory-efficient paging. Training ran for 7 epochs with batch size 1, 8 gradient accumulation steps, and a cosine decay scheduler with 3\% warm-up.

\subsubsection{Evaluation Data} 
Due to the significant class imbalance in the original dataset (15,084 true secrets out of 97,479 candidates), we curated a balanced and representative subset for model training and evaluation. Specifically, we randomly selected 30,000 non-secret samples and retained all 15,000 true secrets, forming a pool of 45,000 labeled examples.

From this curated pool, we created four distinct subsets to support both balanced and imbalanced training scenarios, as well as model evaluation:
\begin{itemize}
    \item \textbf{Balanced Train Set:} 12,000 true secrets and 12,000 non-secrets (24,000 total)
    \item \textbf{Imbalanced Train Set:} 3,750 true secrets and 20,250 non-secrets (24,000 total)
    \item \textbf{Validation Set:} 1,500 true secrets and 1,500 non-secrets (3,000 total)
    \item \textbf{Test Set:} 1,500 true secrets and 1,500 non-secrets (3,000 total, same set as Section \ref{sec:llm_detection})
\end{itemize}

\textbf{Evaluation Metrics: } Similar to Section \ref{sec:llm_detection}, we used those evaluation metrics to evaluate the detected secrets.

\subsubsection{Multiclass classification}
To overcome the limitations of pretrained LLMs, we fine-tuned smaller language models on 9,000 labeled positive samples, specifically curated for the secret classification task. These models were then evaluated on the same 3,000-sample test set used for the pretrained models to ensure a fair comparison.

\section{Result}
This section presents the evaluation results for our two primary research questions. First, we assess the performance of pre-trained large language models (LLMs), such as GPT-4o and DeepSeek-V3, using various prompting strategies. Then, we investigate the effectiveness of smaller, open-source models that have been fine-tuned for secret detection.
\subsection{Secret Detection Using Pre-trained Large Language Models (RQ1)}
To answer RQ1, we evaluate GPT-4o and DeepSeek-V3 using zero-, one-, and few-shot prompting to assess their ability to detect secrets without fine-tuning. Table~\ref{llm-results} shows binary classification results, while Tables~\ref{tab:classification_accuracy_pretrained} and ~\ref{tab:classification_report_gpt4o} present multiclass performance using precision, recall, F1, and F2 scores.
\begin{table}[t]
\centering
\caption{Evaluation Metrics for Zero-shot, One-shot, and Few-shot Learning on Pretrained LLMs}
\label{llm-results}
\begin{tabular}{|p{1cm}|c|p{1cm}|p{0.8cm}|p{0.8cm}|p{1cm}|}
\hline
\textbf{Setting} & \textbf{Model} & \textbf{Precision} & \textbf{Recall} & \textbf{F1-Score} & \textbf{F2-score} \\
\hline
\multirow{2}{*}{Zero-Shot} 
& DeepSeek-V3 & 0.7810 & 0.9033 & 0.8377 & 0.8759 \\
& Gpt-4o      & 0.8233 & 0.8633 & 0.8428 & 0.8550 \\
\hline
\multirow{2}{*}{One-Shot} 
& DeepSeek-V3 & 0.7761 & 0.9913 & 0.8716 & 0.9392 \\
& Gpt-4o      & 0.8623 & 0.9813 & 0.9180 & 0.9549\\
\hline
\multirow{2}{*}{Few-Shot} 
& DeepSeek-V3 & 0.7926 & 0.9887 & 0.8799 & 0.9421\\
& Gpt-4o      & 0.9329 & 0.9465 & 0.9392 & 0.9437\\
\hline
\end{tabular}
\end{table}

The results of DeepSeek-V3 and Gpt-4o, as presented in Table~\ref{llm-results}, show that Gpt-4o consistently outperforms DeepSeek in overall \textbf{Precision}, \textbf{F1-Score}, and \textbf{F2-Score} across zero-shot, one-shot, and few-shot settings. Notably, DeepSeek demonstrates marginally higher \textbf{Recall} in all three settings, highlighting its stronger sensitivity to identifying true positives, albeit at the cost of lower precision.

\begin{itemize}
    \item \textbf{Zero-Shot:} DeepSeek-V3 achieves a Recall of 90.33\% and a higher F2-Score of 87.59\%, compared to Gpt-4o's Recall of 86.33\% and F2-Score of 85.50\%. Although Gpt-4o maintains better Precision (82.33\% vs 78.10\%) and F1-Score (84.28\% vs 83.77\%), DeepSeek's superior Recall gives it the edge in F2-Score.

    \item \textbf{One-Shot:} With the addition of a single labeled example, Gpt-4o reaches 86.23\% Precision, 98.13\% Recall, 91.80\% F1-Score, and a strong 95.49\% F2-Score. Although DeepSeek maintains a slightly higher Recall of 99.13\%, its lower Precision results in a reduced F1 (87.16\%) and F2 (93.92\%) score compared to Gpt-4o.

    \item \textbf{Few-Shot:} Gpt-4o attains its best performance with 93.29\% Precision, 94.65\% Recall, 93.92\% F1-Score, and 94.37\% F2-Score. DeepSeek again shows slightly higher Recall (98.87\%), but falls short in F1 (87.99\%) and F2 (94.21\%) scores. These results confirm Gpt-4o's superior ability to generalize effectively when provided with multiple examples.

\end{itemize}

Overall, Gpt-4o demonstrates stronger performance than DeepSeek across all prompting configurations, making it a more effective and reliable choice for secret detection tasks in source code.

Table~\ref{tab:classification_accuracy_pretrained} shows GPT-4o's classification performance, with a weighted F1-score of 0.8108 and F2-Score of 0.7941. The weighted F1and F2-score was chosen as the evaluation metric to account for class imbalance in the dataset.

\begin{table}[h]
\centering
\caption{Overall Multiclass Classification Accuracy for Pretrained Models}
\label{tab:classification_accuracy_pretrained}
\begin{tabular}{|c|c|c|}
\hline
\textbf{Model} & \textbf{F1-score(weighted)} & \textbf{F2-score(weighted)} \\
\hline
DeepSeek-V3 & 0.7338 & 0.7409\\
GPT-4o & 0.8108 & 0.7941\\
\hline
\end{tabular}
\end{table}

Table~\ref{tab:classification_report_gpt4o} presents the detailed classification report. While classes like \textit{Private Key} and \textit{API Key and Secret} achieve relatively high precision and recall, other categories such as \textit{Generic Secret}, \textit{Other}, and \textit{Username} show poor performance, suggesting that zero-shot or few-shot capabilities are insufficient for nuanced secret-type recognition.

\begin{table}[h]
\caption{Multiclass Classification Report for GPT-4o}
\label{tab:classification_report_gpt4o}
\begin{tabular}{|p{3.5cm}|p{1cm}|p{0.7cm}|p{0.7cm}|p{0.7cm}|}
\hline
\textbf{Category} & \textbf{Precision} & \textbf{Recall} & \textbf{F1-Score} & \textbf{F2-Score}\\
\hline
Private Key & 0.9897 & 0.9411 & 0.9648 & 0.9504 \\
API Key and Secret & 0.7491 & 0.6667 & 0.7055 & 0.6817\\
Authentication Key and Token & 0.9009 & 0.8276 & 0.8627 & 0.8413 \\
Other & 0.2373 & 0.1429 & 0.1783 & 0.1553\\
Generic Secret & 0.0844 & 0.4407 & 0.1417 & 0.2390\\
Database and Server URL & 0.5581 & 0.6857 & 0.6154 & 0.6557 \\
Password & 0.8485 & 0.7179 & 0.7778 & 0.7407 \\
Username & 0.0000 & 0.0000 & 0.0000 & 0.0000\\
\hline
\textbf{Weighted F1-score} & \multicolumn{4}{c|}{0.8108} \\
\hline
\textbf{Weighted F2-score} & \multicolumn{4}{c|}{0.7941} \\
\hline
\end{tabular}
\end{table}

These findings highlight the limitations of relying solely on pretrained models for domain-specific classification tasks like secret type detection, where fine-tuning significantly improves both accuracy and category-level performance.

\begin{tcolorbox}[colback=white, colframe=black, boxrule=1pt, left=0pt, right=0pt, top=0pt, bottom=0pt]
\textbf{RQ1 Summary:} This study evaluates the ability of pre-trained large language models (LLMs), GPT-4o and DeepSeek-V3, to detect secrets in source code without fine-tuning. Using 200-character context windows and structured prompts, models were tested in zero-shot, one-shot, and few-shot settings. GPT-4o consistently outperformed DeepSeek-V3 in Precision and F1-Score, achieving the highest overall F1-score of 93.92\% and F2-score of 94.37\% in the few-shot setting. However, DeepSeek-V3 had higher Recall in all cases and a superior F2-score in the zero-shot scenario. While effective for binary classification, both models struggled with multiclass secret-type classification, highlighting the need for fine-tuning for more nuanced tasks.

\end{tcolorbox}

\subsection{Secret Detection Using Small Large Language Models (RQ2)}
For RQ2, we fine-tune smaller models (LLaMA-3.1, Mistral, CodeLLaMA, Gemma, DeepSeek) on the SecretBench dataset and compare them to their raw versions. Table~\ref{llm-small-results} onward summarizes performance, showing that fine-tuning greatly boosts accuracy, even in low-resource settings.
\begin{table}[h]
\centering
\caption{Evaluation Metrics for Prompting Settings on Raw and Fine-Tuned Models (Zero-, One-, Few-Shot)}
\label{llm-small-results}
\begin{tabular}{|p{0.65cm}|p{1.95cm}|p{0.95cm}|c|c|c|}
\hline
\textbf{Setting} & \textbf{Model} & \textbf{Precision} & \textbf{Recall} & \textbf{F1-Score} & \textbf{F2-Score} \\
\hline
\multirow{5}{*}{\shortstack{Zero-\\Shot\\(Raw)}}
& DeepSeek-7B & 0.4105 & 0.4283 & 0.4054 & 0.4246 \\
& Gemma-7B & 0.5501 & 0.5003 & 0.3353 & 0.5095 \\
& LLaMA-3.1-8B & 0.5412 & 0.5130 & 0.4125 & 0.5184 \\
& Mistral-7B & 0.5033 & 0.5033 & 0.5029 & 0.5033 \\
& CodeLLaMA-7B & 0.5453 & 0.5383 & \textbf{0.5207} & \textbf{0.5397} \\
\hline
\multirow{5}{*}{\shortstack{One-\\Shot\\(Raw)}}
& DeepSeek-7B & 0.6864 & 0.5743 & 0.4990 & 0.5937 \\
& Gemma-7B & 0.6741 & 0.6610 & 0.6545 & 0.6636 \\
& LLaMA-3.1-8B & 0.7622 & 0.6920 & \textbf{0.6699} & \textbf{0.705} \\
& Mistral-7B & 0.6848 & 0.5720 & 0.4949 & 0.5915 \\
& CodeLLaMA-7B & 0.6741 & 0.6610 & 0.6545 & 0.6636 \\
\hline
\multirow{5}{*}{\shortstack{Few-\\Shot\\(Raw)}}
& DeepSeek-7B & 0.5216 & 0.5127 & 0.4563 & 0.5145 \\
& Gemma-7B & 0.7286 & 0.7187 & 0.7156 & 0.7207 \\
& LLaMA-3.1-8B & 0.7913 & 0.7913 & \textbf{0.7913} & \textbf{0.7913} \\
& Mistral-7B & 0.7678 & 0.7630 & 0.7619 & 0.764 \\
& CodeLLaMA-7B & 0.7286 & 0.7187 & 0.7156 & 0.7207 \\
\hline
\multirow{5}{*}{\shortstack{Zero-\\Shot\\(Fine-\\tuned)}}
& DeepSeek-7B & 0.9503 & 0.9473 & 0.9472 & 0.9479 \\
& Gemma-7B & 0.9756 & 0.9753 & 0.9753 & 0.9754 \\
& LLaMA-3.1-8B & 0.9861 & 0.9853 & \textbf{0.9852} & \textbf{0.9855} \\
& Mistral-7B & 0.9815 & 0.9825 & 0.9819 & 0.9823 \\
& CodeLLaMA-7B & 0.9716 & 0.9709 & 0.9707 & 0.971 \\
\hline
\end{tabular}
\end{table}

The performance of all five LLMs—DeepSeek, Gemma, LLaMA-3.1-8B, Mistral, and CodeLLaMA—under different prompting settings without fine-tuning and after fine-tuning is summarized in Table~\ref{llm-small-results}. Across all scenarios, the results clearly show that fine-tuning significantly boosts performance. Among the pre-trained models, LLaMA-3.1 8B consistently achieves the best F1-scores in all prompting conditions.

\begin{itemize}
    \item \textbf{Zero-Shot (Raw):} As seen in Table~\ref{llm-small-results}, model performance is relatively low, with F1-scores ranging from 0.33 to 0.52. CodeLLaMA-7B performs the best in this setting with an F1-score of \textbf{0.5207} and F2-score of \textbf{0.5397}.

    \item \textbf{One-Shot (Raw):} LLaMA-3.1-8B achieves the highest F1-score of \textbf{0.6699} and F2-score \textbf{0.705}, followed by Gemma and CodeLLaMA.

    \item \textbf{Few-Shot (Raw):} With more contextual examples, performance improves further. LLaMA-3.1-8B again leads with F1 and F2-scores of \textbf{0.7913}, followed by Mistral-7B and Gemma-7B.

    \item \textbf{Zero-Shot (Fine-Tuned):} Fine-tuning significantly enhances model effectiveness. LLaMA-3.1-8B attains the top F1-score of \textbf{0.9852} and F2-score of \textbf{0.9855}, followed by Gemma and CodeLLaMA. Even DeepSeek-7B, which had low performance in the raw setting, achieves an impressive \textbf{0.9472} F1-score after fine-tuning.
\end{itemize}

\begin{figure}[t]
\centering
\includegraphics[width=0.4\textwidth]{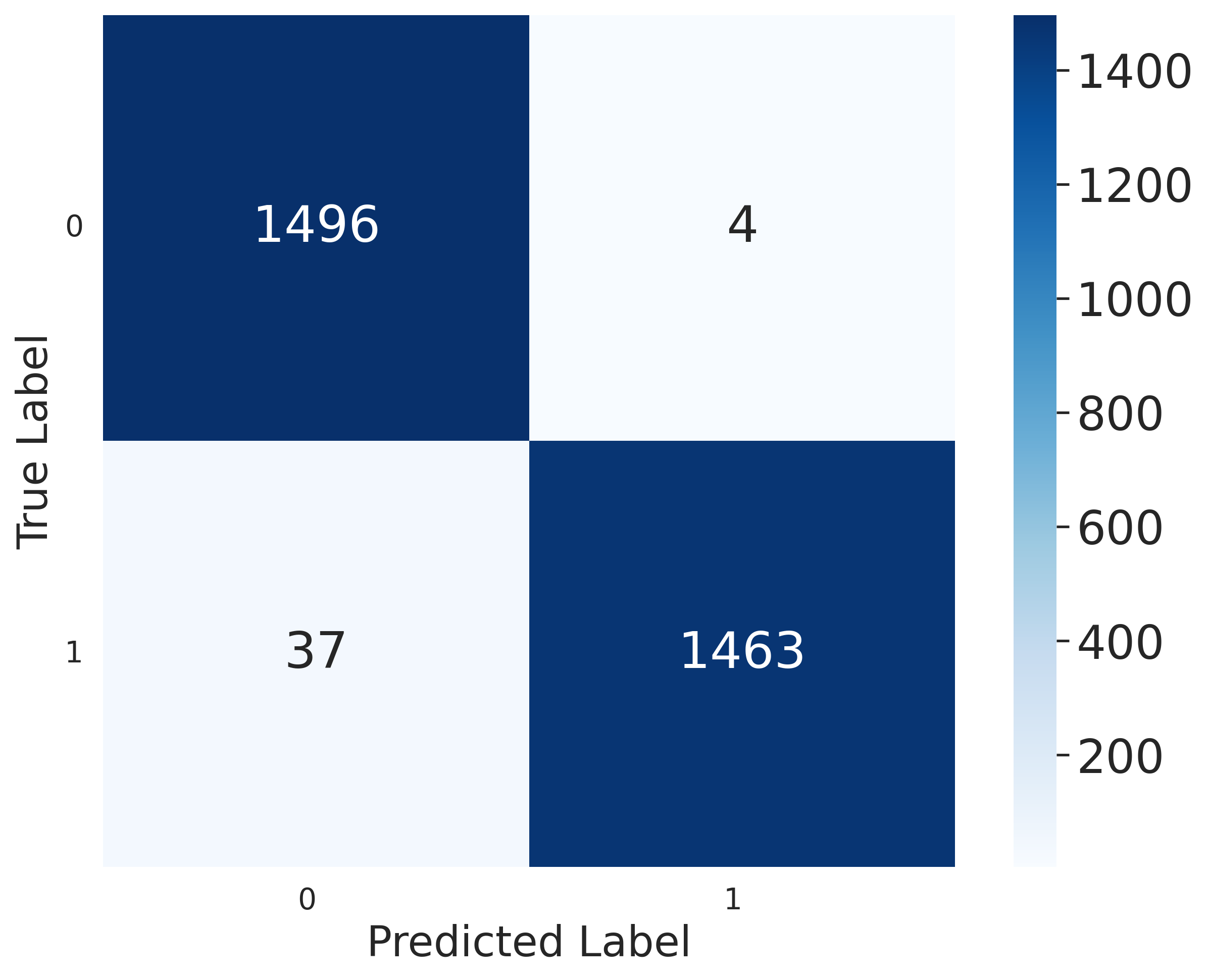}
\caption{Confusion Matrix for Zero-Shot prompting on fine-tuned LLaMA-3.1-8B model}
\label{fig:confusion-matrix-llama3}
\end{figure}

The confusion matrix in Figure~\ref{fig:confusion-matrix-llama3} illustrates the performance of the fine-tuned LLaMA-3.1-8B model under the zero-shot prompting in fine-tuned model. The model correctly identified 1,496 out of 1,500 negative (non-secret) instances and 1,463 out of 1,500 positive (secret) instances. Only 4 false positives and 37 false negatives were observed, indicating high precision and recall. These results are consistent with the high F1-score (0.9852) observed in Table~\ref{llm-small-results}, affirming the model's strong discriminative ability after fine-tuning. The near-diagonal dominance of the matrix further emphasizes the model's robustness in accurately separating secret and non-secret examples.

Overall, these results demonstrate that fine-tuning and prompt-based learning significantly enhance the secret classification capabilities of LLMs, with LLaMA-3.1-8B consistently delivering top-tier performance.
\subsubsection{Multiclass Classification Performance}

Table~\ref{tab:overall_multiclass_accuracy} presents the overall weighted F1-scores for various LLMs fine-tuned for multiclass classification. Among the evaluated models, Mistral-7B achieves the highest weighted F1-score of 0.982, closely followed by LLaMA-3.1-8B and Codellama-7B.

\begin{table}[t]
\centering
\caption{Overall Multiclass Classification Accuracy}
\label{tab:overall_multiclass_accuracy}
\begin{tabular}{|c|c|c|}
\hline
\textbf{Model} & \textbf{F1-Score (Weighted)}& \textbf{F2-Score(Weighted)}\\
\hline
Deepseek-7B & 0.9584 & 0.9581 \\
Gemma-7B & 0.9494 & 0.9472\\
CodeLLama-7B & 0.9625 & 0.9618 \\
LLaMA-3.1-8B & 0.9800 & 0.9798\\
Mistral-7B & \textbf{0.9820} & \textbf{0.9822} \\
\hline
\end{tabular}
\end{table}

\begin{figure}[!h]
    \centering
    \includegraphics[width=1\linewidth]{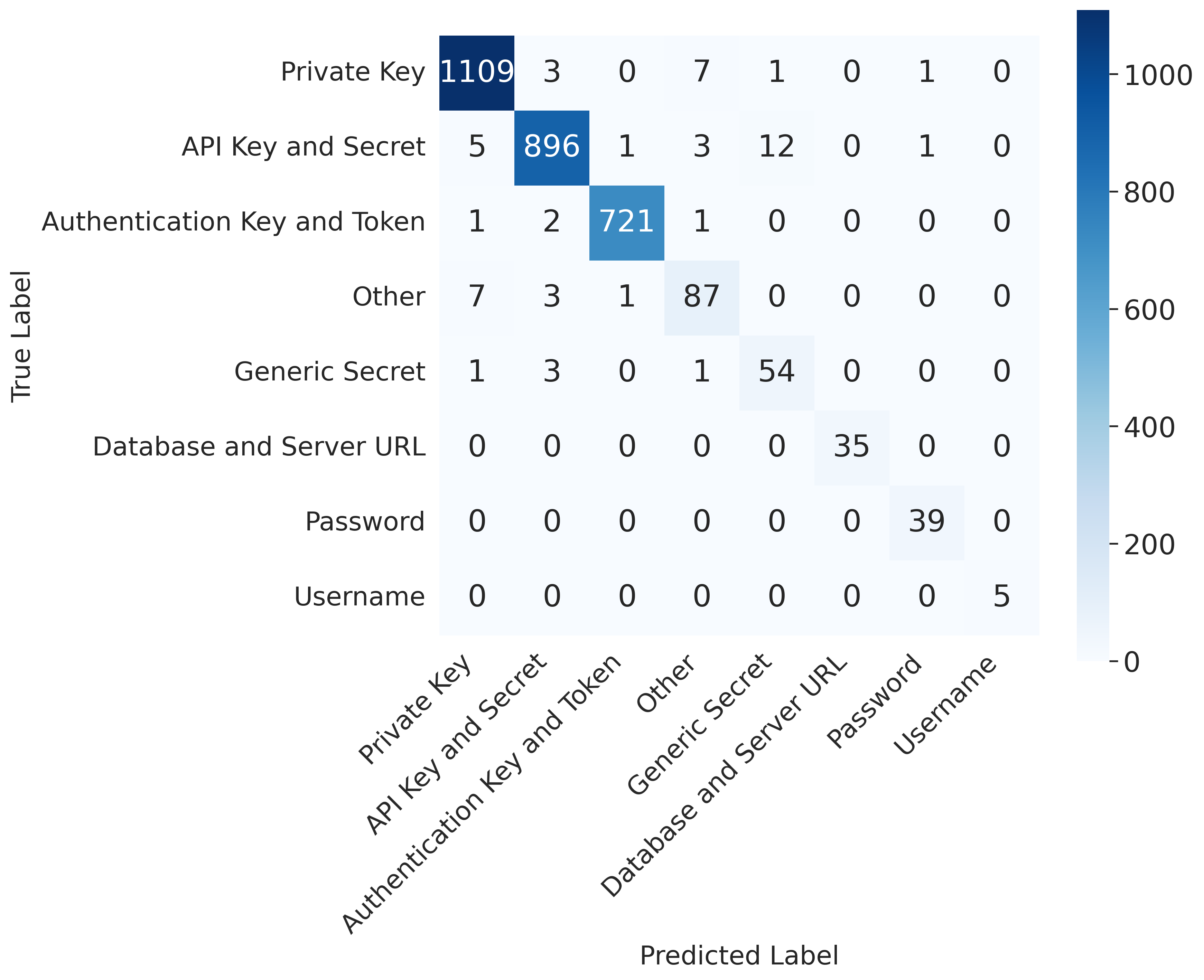}
    \caption{Confusion Matrix for Mistral-7B fine-tuned for multiclass classification}
    \label{fig:conf_matrix_multi}
\end{figure}

To further analyze the best-performing model, Table~\ref{tab:classification_report_mistral} presents the detailed classification report for Mistral-7B across all secret types. Figure~\ref{fig:conf_matrix_multi} shows its confusion matrix, highlighting per-category prediction strengths and weaknesses.

\begin{table}[h]
\centering
\caption{Multiclass Classification Report for Mistral-7B}
\label{tab:classification_report_mistral}
\begin{tabular}{|p{3.5cm}|p{1cm}|p{0.7cm}|p{0.7cm}|p{0.7cm}|}
\hline
\textbf{Category} & \textbf{Precision} & \textbf{Recall} & \textbf{F1-Score} & \textbf{F2-Score} \\
\hline
Private Key & 0.9875 & 0.9893 & 0.9884 & 0.9889 \\
API Key and Secret & 0.9879 & 0.9760 & 0.9819 & 0.9784 \\
Authentication Key and Token & 0.9972 & 0.9945 & 0.9959 & 0.9950 \\
Other & 0.8788 & 0.8878 & 0.8832 & 0.8860 \\
Generic Secret & 0.8060 & 0.9153 & 0.8571 & 0.8911\\
Database and Server URL & 1.0000 & 1.0000 & 1.0000 & 1.0000\\
Password & 0.9512 & 1.0000 & 0.9750 & 0.9898\\
Username & 1.0000 & 1.0000 & 1.0000 & 1.0000\\
\hline
\textbf{Weighted F1-score} & \multicolumn{4}{c|}{0.9820} \\
\hline
\textbf{Weighted F2-score} & \multicolumn{4}{c|}{0.9822} \\
\hline
\end{tabular}
\end{table}

The Mistral-7B model excels in high-frequency categories such as \textit{Private Key} (1,109 correct), \textit{API Key and Secret} (896), and \textit{Authentication Key and Token} (721), showing strong discriminative ability. Some misclassifications occur, particularly between \textit{API Key and Secret}, \textit{Other}, and \textit{Authentication Key and Token}, likely due to feature overlap. Even lower-frequency categories like \textit{Generic Secret}, \textit{Database and Server URL}, and \textit{Password} are accurately classified in most cases.

\begin{tcolorbox}[colback=white, colframe=black, boxrule=1pt, left=0pt, right=0pt, top=0pt, bottom=0pt]
\textbf{RQ2 Summary:} We evaluated the effectiveness of fine-tuned open-source LLMs (e.g., LLaMA-3.1-8B, Mistral, CodeLLaMA, DeepSeek, and Gemma) for secret detection using both binary and multiclass classification. Fine-tuning significantly improved performance, with LLaMA-3.1-8B achieving the highest F1 and F2-score (0.9852 and 0.9855) in zero-shot settings. In multiclass classification, both LLaMA-3.1-8B and Mistral-7B achieved high F1 and F2-scores (98\%+), with Mistral showing strong category-wise precision and recall. These results affirm that compact, fine-tuned models are highly effective and resource-efficient alternatives to commercial LLMs for sensitive software engineering tasks like secret detection.
\end{tcolorbox}

\section{Discussion}
\label{sec:discussion}
\subsection{Analysis of Detected Secrets}

To evaluate the effectiveness of the fine-tuned LLaMA model in binary classification, we analyze its ability to distinguish between true secrets and non-secret content. 

\paragraph{True Positives} The model successfully identified sensitive information, such as an API key embedded in an HTTP request and a cryptographic private key used for decryption. These examples demonstrate the model’s robustness in recognizing structured secrets that adhere to common patterns. For instance, the API key (\texttt{sk\_test\_4eC39HqLyjWDarjtT1zdp7dc}) within a request to Stripe's API was correctly classified as a secret, as was the cryptographic private key in a decryption function. This suggests that the model has effectively learned to detect well-structured secrets present in programming environments.

\begin{lstlisting}[language=Python]
import requests

API_KEY = "sk_test_4eC39HqLyjWDarjtT1zdp7dc"  

response = requests.get(f"https://api.stripe.com/v1/charges", headers={{
    "Authorization": f"Bearer API_KEY"
}})
print(response.json())
\end{lstlisting}

\paragraph{False Positives} The model correctly classified non-secret content as non-sensitive, avoiding false positives in cases where misleading patterns could exist. One such example is a login function containing hardcoded credentials (\texttt{"user123"} and \texttt{"password"}). While these may appear to be sensitive at first glance, they lack complexity and randomness typical of actual secrets. Similarly, an obfuscated API key (\texttt{"xxxxxxxxxx"}) was correctly ignored, likely due to the model recognizing it as a placeholder rather than an actual sensitive credential.
\begin{lstlisting}[language=Python]
def login_user(username, password):
    if username == "user123" and password == "password":
        print("Login successful")
    else:
        print("Login failed")
\end{lstlisting}
\begin{lstlisting}[language=Python]
import requests

API_KEY = "xxxxxxxxxx"  

response = requests.get(f"https://api.stripe.com/v1/charges", headers={{
    "Authorization": f"Bearer API_KEY"
}})
print(response.json())
\end{lstlisting}
These results highlight the model's strong performance in distinguishing true secrets from false alarms. However, future improvements could focus on refining detection criteria to minimize edge cases where synthetic or low-entropy secrets might be misclassified.

\subsection{Comparison with Existing Tools}

Regex-based tools such as \texttt{TruffleHog} are effective for identifying candidate secret strings but suffer from high false positive rates due to a lack of contextual understanding. Since our dataset is constructed from regex-extracted candidates, many of which are false positives, it inherently reflects this limitation. While they exhibit high recall, they suffer from low precision due to an inability to disambiguate true secrets from benign strings.

The best-performing fine-tuned LLaMA-3.1-8B model achieved an F1-score of 0.9852 on our subset of 3,000 samples (1,500 secrets and 1,500 non-secrets). In contrast, a regex-only tool would flag all candidates as secrets, yielding a theoretical precision of just 50\% in this balanced setting.

Our LLM-based classifier thus significantly improves precision while maintaining high recall. By leveraging the contextual understanding of LLMs, we reduce noise introduced by regex and filter out false positives more effectively. This hybrid approach combines the broad detection capability of regex with the discriminative power of LLMs, making it more practical for integration into modern development workflows.

\subsection{False Negatives}

Despite strong performance, our fine-tuned models produced 37 false negatives during evaluation. Manual analysis revealed key causes behind these misclassifications.

One issue stemmed from inconsistencies between our annotation framework and SecretBench’s labeling conventions \cite{basak2023secretbench2}. For example, the dataset labels certain public keys like \texttt{consensus public key} as sensitive, whereas our model treated them as non-sensitive based on our defined classification criteria.. A similar discrepancy was found in cases such as \texttt{pk\_test\_...}, which our model interpreted as a benign public key based on its naming pattern and prior examples. However, the dataset labeled it as sensitive, potentially reflecting real-world risks associated with leaked test tokens, even when such keys are publicly documented to be accessible from the client side (e.g., Stripe documentation). These cases highlight ambiguity in defining sensitive identifiers.

Another factor was the model's limited ability to detect secrets with insufficient context—particularly for long-format secrets like RSA keys, which exceeded the default input size. Increasing the input length from 512 to 1024 tokens improved performance by reducing such false negatives.

A very small number of false negatives arose due to subtle variations in context. Identical candidate strings appeared in different code environments and received inconsistent classifications. This suggests that the model may be sensitive to minor contextual cues, leading to inconsistent decision-making. This represents a potential area for future improvement, particularly in enhancing the model’s robustness to contextual variability.

Overall, the limited false negatives reflect a trade-off favoring lower false positives. By aligning with strict classification criteria, we prioritized precision, which is crucial for maintaining developer trust and usability in practical deployments.

\subsection{Impact of Imbalanced Training Distribution}
Alongside balanced training, we also fine-tuned our models using an imbalanced dataset that mirrors the real-world distribution of secrets in source code. This dataset included 3,750 true secrets and 20,250 non-secret samples, totaling 24,000 instances—closely reflecting the SecretBench ratio, where only 15,084 out of 97,479 samples are actual secrets.

Despite this class imbalance, our models maintained strong performance. As shown in Table~\ref{finetuned-zeroshot-small-imb}, LLaMA-3.1-8B achieved a precision of 0.9742, recall of 0.9730, F1-score of 0.9730 and F2-score of 0.9732 when evaluated on a balanced test set. When trained on the imbalanced dataset, these metrics declined only slightly (by 1–3\%), indicating that LLMs can generalize well even with skewed training distributions.

This robustness can be attributed to the contextual understanding and token-level reasoning inherent in LLMs, which allows them to distinguish secrets from benign strings based on nuanced patterns. Unlike traditional classifiers that often degrade under imbalance, LLMs retain both high recall and precision thanks to their extensive pretraining and semantic awareness.

These findings suggest that LLM-based models remain highly effective even when trained under realistic, imbalanced conditions. This makes them particularly suitable for real-world deployments, where secrets are rare but detecting them accurately is critical for maintaining software security.

\begin{table}[h]
\centering
\caption{Evaluation Metrics for Models Fine-tuned on Imbalanced Dataset}
\label{finetuned-zeroshot-small-imb}
\begin{tabular}{|l|c|c|c|c|}
\hline
\textbf{Model} & \textbf{Precision} & \textbf{Recall} & \textbf{F1-Score} & \textbf{F2-Score} \\
\hline
DeepSeek-7B & 0.9241 & 0.9117 & 0.911 & 0.9142 \\
Gemma-7B & 0.9656 & 0.9451 & 0.9523 & 0.9491 \\
LLaMA-3.1-8B & 0.9742 & 0.973 & \textbf{0.9730} & \textbf{0.9732} \\
Mistral-7B & 0.9715 & 0.9724 & 0.9720 & 0.9722 \\
CodeLLaMA-7B & 0.9571 & 0.9533 & 0.9532 & 0.9541 \\
\hline
\end{tabular}
\end{table}

\subsection{Analysis of Impact of Context Window}

To evaluate the effect of varying the context window, we experimented with two different window sizes: 200 and 300 characters surrounding the candidate secret string. These windows help the model understand the surrounding semantics, which is crucial for accurate classification.

Table~\ref{tab:context-window-comparison} shows the F1-scores obtained by each fine-tuned model when tested under these two context window sizes.

\begin{table}[h]
\centering
\caption{Impact of Context Window Size on F1 and F2 Scores (Fine-Tuned Models)}
\label{tab:context-window-comparison}
\scriptsize
\begin{tabular}{|l|c|c|c|c|}
\hline
\textbf{Model} & \textbf{F1 (200)} & \textbf{F2 (200)} & \textbf{F1 (300)} & \textbf{F2 (300)} \\
\hline
DeepSeek-7B     & 0.9472 & 0.9601 & 0.9690 & 0.9784 \\
Gemma-7B        & 0.9753 & 0.9809 & 0.9764 & 0.9815 \\
LLaMA-3.1-8B    & 0.9852 & 0.9895 & \textbf{0.9882} & \textbf{0.9887} \\
Mistral-7B      & 0.9819 & 0.9870 & 0.9861 & 0.9883 \\
CodeLLaMA-7B    & 0.9707 & 0.9755 & 0.9721 & 0.9767 \\
\hline
\end{tabular}
\end{table}

The results indicate that increasing the context window from 200 to 300 characters leads to a modest improvement in F1 and F2-score across all models. This improvement is attributed to the additional contextual cues available to the model for making more informed classifications—especially useful for complex or multiline secrets.

However, the gains are small, indicating that LLMs, even with smaller context windows, are already powerful enough to perform well due to their pretraining on massive and diverse code/text corpora. Importantly, smaller context windows reduce training and inference times, resulting in lower memory usage and faster model response—an important consideration for deployment in time-sensitive or resource-constrained environments.

Therefore, the choice of context window is a tradeoff between accuracy and resource, with 200 characters already offering strong performance in most use cases.

\subsection{Training Setup}
All fine-tuning experiments were conducted on a local workstation equipped with an Intel i5-13400F CPU, 128 GB RAM, and an NVIDIA RTX 4090 GPU. We utilized QLoRA and PEFT techniques to enable memory-efficient training of 7B–8B models. This setup ensures that the experiments can be reproduced on advanced consumer-grade hardware without requiring access to large-scale cloud infrastructure.

\subsection{Training and Inference Efficiency}
We evaluated the resource usage and performance of each model during both training and inference. Despite variations in model size and architecture, GPU VRAM usage during training remained below 17GB. As shown in Table~\ref{tab:resource-efficiency}, training time ranged from 16 to 25 hours, and inference memory usage stayed below 13GB. Inference latency per sample varied slightly across models, with the fastest being DeepSeek-7B at 0.091 seconds. These results highlight the feasibility of integrating these fine-tuned models into real-time or near-real-time developer environments.

\begin{table}[h]
\centering
\caption{Resource Efficiency Metrics for Fine-Tuned Models}
\label{tab:resource-efficiency}
\scriptsize
\begin{tabular}{|p{1.8cm}|p{0.8cm}|p{0.5cm}|p{1cm}|p{1cm}|}
\hline
\textbf{Model} & \textbf{Train VRAM (GB)} & \textbf{Train Time (hrs)} & \textbf{Inference VRAM (GB)} & \textbf{Inference Time (s)} \\
\hline
DeepSeek-7B     & 9.95 & 16.3  & 10.18 & 0.091 \\
Gemma-7B        & 16.3 & 23.5  & 12.02 & 0.112 \\
LLaMA-3.1-8B    & 15.02 & 24.3 & 9.08 & 0.100 \\
Mistral-7B      & 10.08 & 24.75 & 8.21 & 0.136 \\
CodeLLaMA-7B    & 9.98 & 22.5  & 10.35 & 0.122 \\
\hline
\end{tabular}
\end{table}

\section{Threats to Validity}
\label{sec:threats}
\noindent \textbf{Internal validity} concerns potential issues within the study that could affect the credibility of the results. An internal threat stems from hyperparameter tuning. While we adopted LoRA-based fine-tuning with reasonable settings, we did not perform exhaustive hyperparameter search due to resource constraints. Future experiments could explore further optimization to improve generalization. Additionally, since SecretBench \cite{basak2023secretbench2} is derived from public GitHub repositories, there exists a potential risk of data contamination i.e., LLMs used in our study may have been exposed to similar or identical code during pretraining. However, we posit that even if some of this content was seen during training, the LLMs were unlikely to have been explicitly optimized to detect secrets in those contexts. As such, the learned knowledge is probably not aligned with the secret detection task, and our fine-tuning and prompt-based strategies remain essential for effective performance.

\noindent \textbf{External validity} refers to the generalizability of our results beyond the dataset and models used. Our evaluation focuses on a curated benchmark of GitHub repositories and fine-tuned models. These models may not generalize well to proprietary repositories or other languages and frameworks not represented in our dataset. To reduce this threat, we included diverse programming languages, secret types, and file types, aiming to simulate real-world scenarios. Nonetheless, further testing on industrial codebases would be necessary to validate broader applicability.

\noindent \textbf{Construct validity} addresses whether the study effectively captures what it aims to measure—in this case, true secret detection. One threat is the subjectivity in prompt design and few-shot examples used in LLM prompting. The selection of examples might influence model behavior, particularly for ambiguous candidates. We mitigated this by using representative, diverse examples and evaluating multiple prompting techniques (zero-, one-, and few-shot). Another construct-related concern is the exploitability of detected secrets. Although labeled as true secrets, not all exposed tokens may be exploitable or valid. Due to ethical and legal reasons, we did not test real-world exploitability. This limitation may affect interpretations of the model’s effectiveness in practice.

\noindent \textbf{Conclusion validity} reflects the accuracy of the interpretations made from our experimental results. To ensure robust conclusions, we used standard evaluation metrics—precision, recall, F1 and F2-score along with confusion matrix analysis. Still, model performance can vary across deployment scenarios, especially in noisy or highly imbalanced environments. We attempted to reduce individual evaluation bias by relying on automated classification and consistent test datasets. However, model-specific peculiarities or overfitting to the fine-tuning dataset could skew the results. Future replication studies can help verify our findings across alternate settings.

\section{Conclusion}

\label{sec:conclusion}
In this study, we investigated the use of large language models (LLMs) for detecting secrets in source code, addressing the limitations of traditional regex and entropy-based tools, which often suffer from high false positive rates. We introduced a hybrid approach that first uses regex-based candidate extraction and then applies LLM-based classification to filter out false positives. Our results show that fine-tuned models such as LLaMA achieve near-perfect performance in both binary and multiclass classification tasks, while pre-trained models like GPT-4o also perform well in few-shot settings, demonstrating the flexibility of LLMs in security-related applications.

To support our experiments, we curated and balanced a subset from the SecretBench dataset, extracted contextual code snippets, and designed structured prompts for training and inference. Our extensive evaluation across prompting strategies and model types highlights the strong performance of fine-tuned open-source models, which offer practical benefits in environments with privacy constraints or limited access to commercial APIs.

Looking ahead, we plan to integrate our approach into CI/CD pipelines and IDEs for real-time secret detection. Future work includes exploring ensemble models, uncertainty-based rejection to reduce false negatives, and extending detection to other artifacts. We also aim to enhance generalization via domain adaptation and improve adaptive prompting for security-focused tasks.

\bibliographystyle{IEEEtran}
\bibliography{references}

\end{document}